\documentstyle[epsfig,11pt,newpasp,twoside]{article}
\index{G11.2-0.3}
\index{PSR J1811-1925}
\index{supernova remnant}
\index{pulsar wind nebula}
\index{X-ray pulsar}

\def\edcomment#1{\iffalse\marginpar{\raggedright\sl#1\/}\else\relax\fi}
\reversemarginpar

\begin{document}
\title{The Pulsar Wind Nebula of G11.2$-$0.3}
\author{M. S. E. Roberts \altaffilmark{1}, C. Tam, V. M. Kaspi
\altaffilmark{1}, M. Lyutikov \altaffilmark{1}}
\affil{McGill University, 3600 University St., Montreal, QC H3A 2T8,
Canada}
\author {E. V. Gotthelf} 
\affil{Columbia University, 550 West 120th St., New York, NY 10027,
USA}
\author{G. Vasisht}
\affil{Jet Propulsion Laboratory, Caltech, 4800 Oak Grove Dr.,
Pasedena, CA 91109, USA} 
\author{N. Kawai}
\affil{RIKEN, 2-1 Hirosawa, Wako, 351-0198 Saitama, Japan}

\altaffiltext{1}{Department of Physics and Center for Space Research,
MIT, Cambridge, MA 02139, USA}

\begin{abstract}
We present high-resolution radio and X-ray studies of the
composite supernova remnant G11.2$-$0.3.  Using archival VLA data, we
perform radio spectral tomography to measure for the first time the
spectrum of the shell and plerion separately.  We compare the radio
morphology of each component to that observed in the hard and soft
Chandra X-ray images. We measure the X-ray spectra of the shell and
the emission in the interior and discuss the hypothesis that soft
X-ray emission interior to the shell is the result of the expanding
pulsar wind shocking with the supernova ejecta.  We also see evidence
for spatial variability in the hard X-ray emission near the pulsar,
which we discuss in terms of ion mediated relativistic shocks.
\end{abstract} 

\section{Introduction}

The supernova remnant G11.2$-$0.3 is a bright, circular X-ray and
radio shell.  At its center is the X-ray pulsar PSR J1811$-$1925
($P=65$~ms, spin-down energy $\dot E = 6.4 \times 10^{36}$~erg/s,
characteristic age $\tau=24,000$~yrs, Torii et al. 1999) and its
associated hard X-ray wind nebula (Vasisht et al. 1996).  The
characteristic age is much greater than the apparent age of the SNR
($\sim 2000$~yrs), the age implied by its highly centralized position
within the remnant, and its likely association with the historical
event of 386~A.D. (Kaspi et al. 2001).  This discrepancy suggests the
pulsar's current spin period is very near its initial value, and that
$\dot E$ has remained nearly constant since the supernova explosion. 

\section{Radio and X-ray Analysis}

Radio observations of G11.2$-$0.3 were made with the Very Large Array
(VLA) at 20 and 6~cm between April 1984 and May 1985.  To measure the
spectral index $\alpha$, where $S_\nu \propto \nu^{-\alpha}$, we 
spacially filtered the $u-v$ coverages to match spatial scales
(Gaensler et al. 1999) by creating a model visibility data set from
the 20~cm image with the $u-v$ sky distribution of the 6~cm data. A
20~cm image was then created from this model with identical parameters
as the 6~cm image, and both were convolved with a 10$''$ Gaussian. 
\begin{figure}[ht]
\begin{center}
\epsfig{file=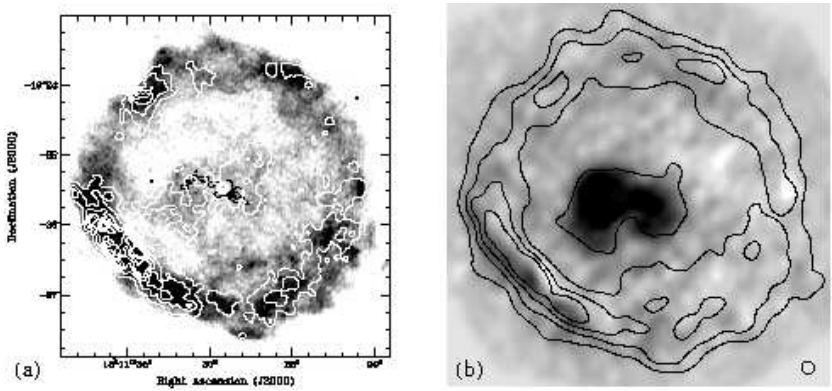,width=11cm}
\caption{(a) High resolution image of G11.2$-$0.3 at 20~cm, with hard
(black) and soft (white) X-ray contour overlays.  (b) Tomographic
difference image for $\alpha_t = 0.56$ at 10$''$ resolution with 6~cm
contours.}
\end{center} 
\end{figure}
To determine spectral indices, we used ``spectral tomography'' 
(Katz-Stone \& Rudnick 1997).  A difference image $I_{\alpha_t}$ was
calculated by scaling the 6~cm image by a trial spectral index
$\alpha_t$, and subtracting it from the 20~cm image:
$I_{\alpha_t}=I_{20}-(\nu_{20}/\nu_6)^{\alpha_t}I_6$, 
where $I_{20}$ and $I_6$ were the images being compared, and
$\nu_{20}$ and $\nu_6$ were the average frequencies of the 
images.  When the trial spectral index reached the actual spectral
index of a particular feature, ie. $\alpha_t = \alpha$, that region
appeared to vanish into the local background of the difference image
(Figure 1).  By examining the resulting series of difference images,
we estimate $\alpha_{P} = 0.25 \pm 0.05$ for the PWN region in the
center, while for the majority of the SNR shell we estimate a value of
$\alpha_{S} = 0.56 \pm 0.02$, which is in agreement with single-dish
radio results as published by Kothes \& Reich (2001).  However, the
SNR shell appears to have a non-uniform spectral index, most
noticeable in the south-eastern region.  Negative residuals can be
seen at $\alpha_t = 0.52$ in the outer component of the SNR shell, and
lingering positive residuals appear at $\alpha_t = 0.58$ in the inner
shell, indicating that the spectrum of the outer shell is slightly
flatter, and the inner shell slightly steeper, than the bulk of the
remnant.

Chandra observed G11.2$-$0.3 with the ACIS-S3 CCD on 2000 August~6
(obs 1) for 20~ks and 2000 October~15 (obs 2) for 15~ks.  The analysis
reported here was performed without the improved S3 FEF files, and we
are currently re-analysing the data using the newer calibration.  We
created exposure corrected ``soft" band (0.6-1.65~keV) and ``hard"
band (4-9~keV) images to show regions of thermal and non-thermal
emission, respectively.  We extracted spectra from the regions
indicated in Figure 2b. The interior background region was used when
fitting features inside the shell and an exterior region as background
for the shell (not shown).

\begin{figure}[h!]
\begin{center}
\epsfig{file=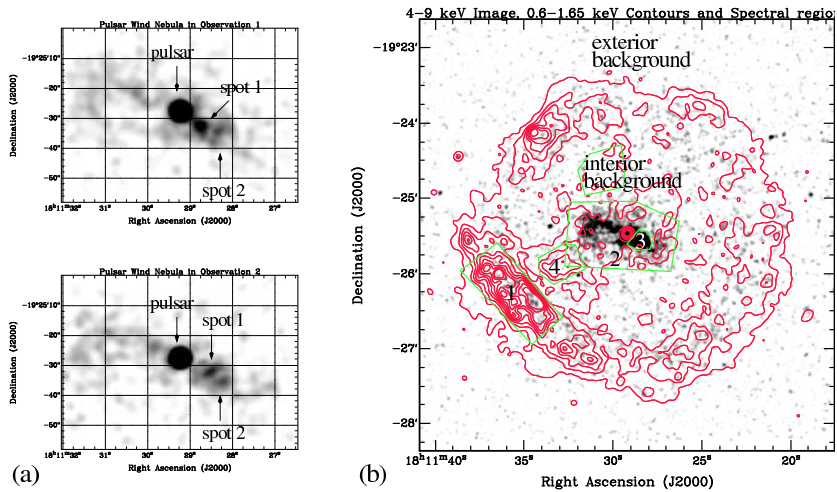,width=11cm}
\caption{(a) Hard X-ray emission of PWN at epoch 1 (top) and epoch 2
(bottom).  (b) Hard X-ray image with soft X-ray contours and spectral
region designations.}
\end{center}
\end{figure}
We fit the shell region (1) and the soft ``PWN" region (4) spectra 
to the VPShock model plus absorption in XSPEC using spatially weighted 
response files, fitting for elemental abundances that could be
constrained by the data. We added a power-law model when fitting
region (1) due to an excess of hard photons in that region.  The
results suggest the region (4) spectrum is slightly cooler than region
(1) ($kT\sim 0.42$~keV vs. $kT\sim 0.55$~keV) and has higher
abundances of O, Ne, Mg, and S.  For the hard PWN (regions 2 and 3),
we fixed the VPShock values from the region (4) fit, allowing only the
normalization to vary and adding a power-law.  The resulting photon
spectral index was $\sim 1.5 \pm 0.2$ and the unabsorbed power-law
flux of regions (2) (the entire PWN minus the pulsar) and (3) (the
bright portion of the PWN) were $\sim 4.1 \times 10^{-12}\,{\rm
erg}\,{\rm cm^{-2}}\,{\rm s^{-1}}$ and $7.3 \times 10^{-13}\,{\rm
erg}\,{\rm cm^{-2}}\,{\rm s^{-1}}$ respectively.

The hard band images from the two observation epochs show two bright
``spots'' in region (3) near the pulsar (Figure 2a).  Assuming a
distance of 5~kpc to the remnant, spot 1 is $\sim 0.20$~pc and spot 2
is $\sim 0.34$~pc from the pulsar. Between epochs, the centroided
positions of the spots moved significantly away from the pulsar,
implying space velocities of $\sim c$, with the relative velocity of
spot 2 $\sim \twothirds$ that of spot 1.

\section{Discussion}

The separation between the PWN and the inner edge of the shell
in both radio and X-rays suggests that the remnant may not have
reached the Sedov phase. This would imply the supernova ejecta within
the shell, which interacts with the PWN, is still in the free
expansion phase. The shell itself, however, has probably swept up
enough mass to be between the free expansion and Sedov solutions.  The
expansion of the PWN into the ejecta is supersonic, and should set up
a forward shock. This heats the ejecta, possibly resulting in the
regions of thermal emission seen interior to the shell. The apparent 
symmetry around the pulsar of this soft ``PWN" emission (region 4),
along with the relatively greater abundances of heavy elements
compared with the bright shell, support this picture. The radio PWN on
the one side is sandwiched between the bright, narrow, jet-like hard
X-ray feature and the soft ``PWN" region, which also suggests a
connection.  Simple spherical models of PWN expansion (eg. Reynolds \&
Chevalier 1984) into the freely expanding interior of an SNR suggest
that the ratio of the PWN radius to the SNR radius should be $\sim
0.2$, assuming that the shell is also freely expanding into the
ISM. The larger observed ratio of $\sim 0.3$ is expected if the shell
has slowed due to accumulated ISM mass but the reverse shock has not
yet encountered the PWN and compressed it. 

The bright spots in the hard X-ray nebula may be the equivalent of the
Crab's ``wisps".  The motion of wisps may be interpreted in the
framework of ion mediated relativistic shocks (Gallant \& Arons 1994). 
We can estimate $\sigma$ (Kennel \& Coroniti 1984), the ratio of
Poynting to particle momentum flux in the wind from the expansion
velocity of the PWN (assuming $d\sim 5$~kpc and an age of 1600~yrs)
as $\sigma \sim V/c \sim 0.002$.  We estimate the magnetic field in
the shocked flow $B \simeq 3 (\sigma \dot{E} /( 4 \pi r_s^2 c))^{1/2}
= 3 \mu G$ if we assume the position of the wind termination shock
$r_s$ is approximately that of spot 1.  If we identify the distance
between the two bright spots, with the ion gyration radius $r_{L,i}$,
we find the ions have passed through $\sim  3 \sigma^{1/2} r_{L,i} /
r_s \sim 0.1$ of the total potential drop through the open
magnetosphere, similar to the Crab's value of 0.3.  We conclude that
both the geometry and dynamics (nearly relativistic motion) of the
bright spots observed near PSR J1811$-$1925 are consistent with the
ion mediated relativistic shock model of Gallant and Arons.

\end{document}